\documentclass[twocolumn,conference]{IEEEtran}
\usepackage[T1]{fontenc}
\usepackage{color}
\usepackage{verbatim}
\usepackage{float}
\usepackage{amsmath}
\usepackage{amsthm}
\usepackage{amssymb}
\usepackage{graphicx}

\makeatletter

\floatstyle{ruled}
\newfloat{algorithm}{tbp}{loa}
\providecommand{\algorithmname}{Algorithm}
\floatname{algorithm}{\protect\algorithmname}

\theoremstyle{plain}
\ifx\thechapter\undefined
\newtheorem{thm}{\protect\theoremname}
\else
\newtheorem{thm}{\protect\theoremname}[chapter]
\fi
\theoremstyle{definition}
\newtheorem{defn}[thm]{\protect\definitionname}

\usepackage[caption=false,font=footnotesize]{subfig}
\usepackage[english]{babel}
\usepackage{times}
\usepackage{color}
\usepackage{amsmath}
\usepackage{amsthm}
\usepackage{amstext}
\usepackage{hyperref} 
\usepackage{amsfonts}
\usepackage{amssymb}
\usepackage{psfrag}
\usepackage{fancyhdr}
 \usepackage{algorithmic}
\allowdisplaybreaks

\IEEEpubidadjcol  \IEEEoverridecommandlockouts
\IEEEpubid{\makebox[\columnwidth]{978--1--5386--3873--6/17/\$31.00~\copyright~2017 IEEE\hfill} \hspace{\columnsep}\makebox[\columnwidth]{ }}

\makeatother

\providecommand{\definitionname}{Definition}
\providecommand{\theoremname}{Theorem}

\begin{document}

\title{Proactive Edge Computing in Latency-Constrained Fog Networks \thanks{ This research was supported by the Academy of Finland (CARMA) project, NOKIA donation on fog (FOGGY project), and by the U.S. Office of Naval Research (ONR) under Grant N00014-15-1-2709. \vspace{-2mm}} \vspace{-0.45cm}
}

\author{\IEEEauthorblockN{Mohammed~S.~Elbamby\IEEEauthorrefmark{1}, Mehdi~Bennis\IEEEauthorrefmark{1}\IEEEauthorrefmark{2},
and Walid~Saad\IEEEauthorrefmark{3}}\IEEEauthorblockA{\IEEEauthorrefmark{1}Centre for Wireless Communications, University
of Oulu, Finland, \\
 emails: \{mohammed.elbamby,mehdi.bennis\}@oulu.fi \\
}\IEEEauthorblockA{\IEEEauthorrefmark{2}Department of Computer Engineering, Kyung Hee
University, South Korea\\
}\IEEEauthorblockA{\IEEEauthorrefmark{3}Wireless@VT, Bradley Department of Electrical
and Computer Engineering, \\
Virginia Tech, Blacksburg, VA, USA, email: walids@vt.edu \vspace{-0.6cm}
}}
\maketitle
\begin{abstract}
In this paper, the fundamental problem of distribution and proactive
caching of computing tasks in fog networks is studied under latency
and reliability constraints. In the proposed scenario, computing can
be executed either locally at the user device or offloaded to an edge
cloudlet. Moreover, cloudlets exploit both their computing and storage
capabilities by proactively caching popular task computation results
to minimize computing latency. To this end, a clustering method to
group spatially proximate user devices with mutual task popularity
interests and their serving cloudlets is proposed. Then, cloudlets
can proactively cache the popular tasks' computations of their cluster
members to minimize computing latency. Additionally, the problem of
distributing tasks to cloudlets is formulated as a matching game in
which a cost function of computing delay is minimized under latency
and reliability constraints. Simulation results show that the proposed
scheme guarantees reliable computations with bounded latency and achieves
up to $91\%$ decrease in computing latency as compared to baseline
schemes.

\vspace{-0.07cm}
\end{abstract}

\section{Introduction}

\vspace{-0.1cm}

The emergence of the Internet of things (IoT) and machine-to-machine
communication is paving the way for a seamless connectivity of a massive
number of resource-limited devices and sensors \cite{walid_mtc}.
The unprecedented amount of IoT data communication and computation
requirements impose stringent requirements in end-to-end latency,
mandating ultra-reliable and low-latency communications (URLLC). However,
the finite computation capabilities of end-user devices challenge
the possibility of coping with the stringent computing and processing
latency requirements of IoT network\textcolor{black}{s. Therefore,
mobile cloud computing (MCC) services have been recently proposed
to allow end-users to offload their resource consuming tasks to remote
cloud centers. However, despite having high computational resou}rces,
MCC solutions are inefficient in handling latency-critical computing
services due to the high propagation delays between the end-user device
and the cloud data center. 

Recently, the idea of fog computing has been introduced \cite{fog_its_role}
to bring computing resources closer to where tasks are requested.\textcolor{black}{{}
In order to minimize computing latency in fog networks, smarter communication
and computing resource utilization schemes are needed \cite{MEC_survey}.
A centralized joint communication and computation resource allocation
sc}heme is proposed in \cite{barbarossa_central_RA}. The power-delay
tradeoff in centralized mobile edge computing (MEC) systems is discussed
in \cite{P_D_tradeoff_letaief} using tools from stochastic optimization.
However, these works rely on centralized solutions in which the MEC
network has information about all users requests and channel-state
information (CSI). Game-theoretic solutions are studied in \cite{decent_game}
to design decentralized computing offloading schemes in cases of homogeneous
and heterogeneous users. Recently, an online secretary framework for
fog network formation is proposed in \cite{secretary} under uncertainty
on the arrival process of fog nodes.

While interesting, the vast majority of the literature in fog networking
is based on the reactive computing paradigm in which task computing
starts only after the task data is offloaded to the fog node\textcolor{black}{{}
\cite{MEC_survey}}. Moreover, prior art has not explicitly accounted
for stringent latency and reliability constraints in fog networks.
Due to the distributed nature of these networks, having computing
resources closer to the network edge allows for providing personalized
type of computing services to end-users \cite{MEC_survey}. Clearly,
harnessing the correlation between end-user requests motivates the
need for \emph{proactive computing} to minimize computing latency.
For example, for augmented reality (AR) services provided in a museum,
proactively computing popular AR services of visitors can aid in minimizing
computational latency \cite{ejder_VR}. Using proactive computing,
the fog network can keep track of the popularity patterns of user
tasks and cache their computing results \emph{in advance}. This eliminates
the need to request the task data multiple times thus reducing the
burden on the task offloading transmissions\textcolor{black}{{} \cite{MEC_survey}.
A possible first step towards proactive computing is the idea of task
data prefetching \cite{prefetching_paper}, in which} part of the
upcoming task data is predicted and prefetched during the computing
of the current one such that the fetching time is minimized. 

While the idea of proactive networks has been recently studied in
the context of wireless content caching, such as \cite{ejder_edge}
and \cite{elbamby_caching}, none of these works investigate the problem
of proactive caching of computing tasks. In contrast to content caching,
computing caching poses new challenges. First, while in content caching,
popular contents are prefetched from the core network during off-peak
times to alleviate the burden on the backhaul links, computing caching
decreases the load on the access link by providing computing results
to end-user nodes (UNs) without the need to prefetch their task data
beforehand. Second, computing tasks can be of diverse types and depend
on the computing environment, while some of the content is cacheable
for reuse by other devices, personal computing data is not cacheable,
and must often be computed in real time. Finally, due to the nature
of IoT networks, with large number of deployed servers and low density
of UNs per server \cite{small_cell_book}, it is not practical to
build popularity patterns locally at each server. Instead, studying
popularity distributions over larger sets of servers can provide a
broader view on the popularity patterns of computing tasks.

\begin{figure}[t]
\includegraphics[width=1\columnwidth]{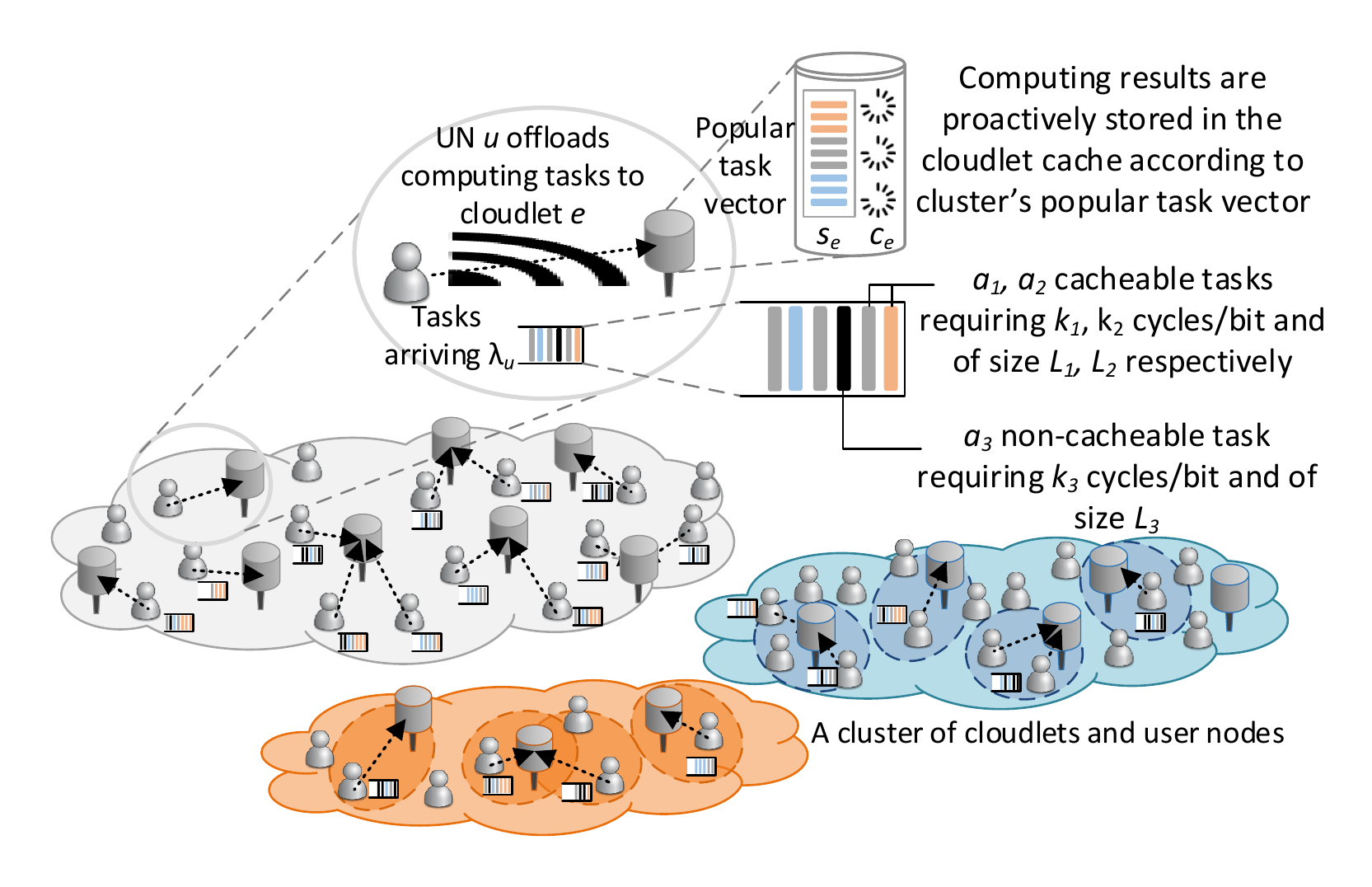}
\centering{}\caption{\label{fig:Fog_net}An illustration of the cache-enabled fog network
model}
\end{figure}

The main contribution of this paper is to investigate the problem
of edge computing and proactive edge caching in fog computing networks.
We exploit both computing and storage resources to minimize computing
latency via joint task offloading and proactive caching of \emph{popular}
and \emph{cacheable} computing tasks. A cacheable task has a computing
result that can be reused by several other devices. Moreover, we impose
constraints on computing latency and reliability so as to ensure computing
delay bounds with high levels of reliability. In the proposed framework,
clusters of UNs and their serving edge computing nodes (cloudlets)
are formed based on spatial proximity and mutual interests in popular
tasks. Accordingly, cloudlets proactively cache computed results of
popular tasks in their clusters thereby ensuring minimal latency.
Moreover, the problem of task distribution to cloudlets is modeled
as a matching game between cloudlets and UNs. To this end, an efficient
distributed matching algorithm is proposed to reach a stable matching
of UN requests to cloudlets or to their local device such that the
minimal task computing delay is incurred and a reliable service latency
is guaranteed. Simulation results show that the proposed framework
can guarantee reliable computations with bounded latency and achieve
up to $91\%$ decrease in computing latency as compared to baseline
schemes.

The rest of this paper is organized as follows. Section~\ref{sec:SysMod}
describes the system model and problem formulation. The proposed clustering
scheme as well as the joint caching and matching scheme are studied
in Section~\ref{sec:JointTandM}. The performance of the proposed
framework is analyzed in Section~\ref{sec:SimRes}. Finally, Section~\ref{sec:Conc}
concludes the paper.

\section{System Model\label{sec:SysMod}}

\vspace{-0.1cm}

Consider a fog network that consists of a set $\mathcal{E}$ of $E$
cloudlets, each of which having a CPU computing and storage capability
of $c_{e}$ and $s_{e}$, respectively, and a set \textbf{$\mathcal{U}$}
of $U$ UNs that are distributed uniformly over the network are\textcolor{black}{a.
Cloudlets share the same frequency channel and operate in time-division-duplex
(TDD). In our model, we focus on the uplink transmission. U}Ns have
computing tasks that arrive following a Poisson process with mean
$\lambda_{u}$. UNs are interested in a set $\mathcal{A}$ of $A$
tasks. Each task has a required  CPU cycles of $\kappa$ per bit of
task data\textcolor{black}{, and the task data size follows an exponential
distribution of mean $L_{a}$. UNs can offload their computing tasks
to any cloudlet within their coverage, where coverage is decided based
on a threshold path loss value}\footnote{\textcolor{black}{Path loss is used as a coverage metric such that
UN's cloudlet list does not change frequently due to wireless channel
dynamics.}}\textcolor{black}{. Task data must be offloaded to} the cloudlet prior
to computation. To minimize the task data offloading delay, cloudlets
proactively cache the computed results of the most popular cacheable
tasks. We assume that a subset $\mathcal{A}_{c}\subset\mathcal{A}$
of the tasks are cacheable and another subset $\mathcal{A}_{nc}\subset\mathcal{A}$
is non-cacheable such that $A_{c}\cup\mathcal{A}_{nc}=\mathcal{A}$.
An illustration of the studied fog network model is shown in Fig.~\ref{fig:Fog_net}.

\vspace{-0.1cm}

\subsection{Computing Model}

\vspace{-0.1cm}

The computation of each task $a\in\mathcal{A}$ by UN $u$ can be
either performed locally or offloaded to a cloudlet. The total local
computing time is: 

\vspace{-0.2cm}

\begin{equation}
D_{ua}^{\textrm{l}}(t)=\frac{\kappa L_{a}}{c_{\textrm{local}}}+W_{ua}^{\textrm{l}}(t)+\text{\ensuremath{\tau}}_{\textrm{LP}},
\end{equation}
where $W_{ua}^{\textrm{l}}(t)$ is the queuing delay of task $a$
in the local queue of UN $u$ at time $t$, $c_{\textrm{local}}$
is the local computing capability in cycles/second, and $\text{\ensuremath{\tau}}_{\textrm{LP}}$
is the local processing delay. 

Each task $a$ requested by user $u$ and offloaded to the cloudlet
$e$ experiences a total computing delay that consists of the task
data transmission time, cloudlet computing time, cloudlet queuing
time, and processing time, as follows:\vspace{-0.3cm}

\begin{equation}
D_{ea}^{\textrm{f}}(t)=\biggl(\frac{\kappa L_{a}}{c_{e}}+\frac{L_{a}}{r_{ue}(t)}+W_{ea}^{\textrm{f}}(t)\biggr)(1-y_{ea}(t))+\text{\ensuremath{\tau}}_{\textrm{EP}},
\end{equation}

\vspace{-0.2cm}
{\setlength{\parindent}{0cm}\textcolor{black}{where $r_{ue}=\textrm{BW}\log_{2}\Bigl(1+P_{u}h_{ue}/(N_{o}+I_{e})\Bigr)$
is the uplink data rate}\footnote{BW\textcolor{black}{{} is the channel bandwidth, $P_{u}$ is the transmit
power of UN $u$, $h_{ue}$ is the channel gain between UN $u$ and
cloudlet $e$, $I_{e}$ is the interfering power from other UNs, and
$N_{o}$ is the noise power. }}\textcolor{black}{{} from UN $u$ to the cloudlet $e$, $W_{ea}^{f}(t)$
is the waiting time of task $a$ due to the previous computing tasks
in the queue $Q_{e}$ of cloudlet $e$, $y_{ea}(t)$ is a binar}y
variable that equals $1$ when the computation result of task $a$
is cached in cloudlet $e$, and $\text{\ensuremath{\tau}}_{\textrm{EP}}$
is the cloudlet latency which accounts for the downlink transmission
of computed data and the cloudlet processing latency. Consequently.
the delay incurred by the computation of a given task $a$ will be:}

\vspace{-0.5cm}

\begin{equation}
D_{a}(t)=x_{ea}(t)D_{ea}^{\textrm{f}}(t)+\biggl(1-\sum_{e\in\mathcal{E}}x_{ea}(t)\biggr)D_{ua}^{\textrm{l}}(t),
\end{equation}

\vspace{-0.2cm}
{\setlength{\parindent}{0cm}where $x_{ea}$ is a binary variable
that equals $1$ if task $a$ is distributed to cloudlet $e$. }

Similar to other works \cite{prefetching_paper}, we assume that the
latency due to downlink transmission of computed data is negligible
compared to the uplink task data offloading time and computing time,
and, hence, it is not accounted for in the optimization problem. This
assumption is due to the typically small size of computed data and
the relatively high transmission power of cloudlet compared to end-user
devices. 

Our objective is to minimize the total task computing latency under
reliability constraints, by efficiently distributing and proactively
caching the results of computing tasks. The UN task distribution to
cloudlets and task caching matrices are expressed as $\boldsymbol{X}=[x_{ea}]$
and $\boldsymbol{Y}=[y_{ea}]$, resp\textcolor{black}{ectively. Reliability
is modeled as a probabilistic constraint on the maximum offloaded
computing delay. This op}timization problem is: 

\vspace{-0.5cm}

\begin{subequations}\label{eq:objective} 

\begin{align}
\min_{\boldsymbol{X},\boldsymbol{Y}} & \sum_{u\in\mathcal{U}}D_{a}(t)\\
 & \Pr(D_{ea}^{\textrm{f}}(t)\geq D_{\textrm{th}})\leq\epsilon,\;\forall e\in\mathcal{E},\label{eq:prob_const}\\
 & \sum_{e\in\mathcal{E}}x_{ea}(t)\leq1,\;\forall u\in\mathcal{U},\label{eq:match_const_1}\\
 & \sum_{a\in Q_{e}}x_{ea}(t)\leq1,\;\forall e\in\mathcal{E},\label{eq:match_constraint_2}\\
 & \sum_{a\in\mathcal{A}}y_{ea}(t)\leq s_{e},\;\forall e\in\mathcal{E},\label{eq:cache_const}
\end{align}
\end{subequations}where (\ref{eq:prob_const}) is a probabilistic
delay constraint that ensures the latency is bounded by a threshold
value $D_{\textrm{th}}$ with a probability $1-\epsilon$. Constraints
(\ref{eq:match_const_1}) and (\ref{eq:match_constraint_2}) ensure
the one-to-one correspondence of distributing new requests to cloudlets.
(\ref{eq:cache_const}) limits the number of cached tasks to a maximum
of $s_{e}$. The above problem is a combinatorial problem with a non-convex
cost function and probabilistic constraints, for which finding an
optimal solution is computationally complex \cite{QoS_const}. The
non-convexity is due to the service rate term in the delay equation
which is function of the interference from other offloading UNs. To
make the problem tractable, we use the Markov's inequality to convert
the probabilistic constraint in (\ref{eq:prob_const}) to a linear
constraint \cite{QoS_const} expressed as $\mathbb{E}\{D_{ea}^{\textrm{f}}(t)\}\leq D_{\textrm{th}}\epsilon,$
where $\mathbb{E}\{.\}$ denotes the expectation over time. Since
the delay of computing a cached task is very small, we are interested
in keeping the delay of non-cached tasks below a pre-defined threshold.
Hence, the constraint can be written as:

\vspace{-0.3cm}

\begin{equation}
\mathbb{E}\biggl\{\frac{\kappa L_{a}}{c_{e}}+\frac{L_{a}}{r_{ue}(t)}+W_{ea}^{\textrm{f}}(t)+\text{\ensuremath{\tau}}_{\textrm{EP}}\biggr\}\leq D_{\textrm{th}}\epsilon,
\end{equation}
\vspace{-0.3cm}
substituting the queuing time as $W_{ea}^{\textrm{f}}(t)=\sum_{a_{i}\in Q_{e}}\frac{L_{a_{i}}'(t)}{r_{ie}(t)}$:

\begin{equation}
\mathbb{E}\biggl\{\frac{L_{a}}{r_{ue}(t)}\biggr\}\leq D_{\textrm{th}}\epsilon-\mathbb{E}\biggl\{\sum_{a_{i}\in Q_{e}}\frac{L_{a_{i}}'(t)}{r_{ie}(t)}\biggr\}-\frac{\kappa L_{a}}{c_{e}}-\text{\ensuremath{\tau}}_{\textrm{EP}},
\end{equation}
where $L_{a_{i}}'(t)$ is the remaining task data of task $a_{i}$
in the queue $Q_{e}$ of cloudlet $e$ at time instant $t$. Finally,
the constraint can be expressed as:\vspace{-0.3cm}

\begin{equation}
\frac{L_{a}}{\bar{r}_{ue}(t)}\leq D_{\textrm{th}}\epsilon-\frac{\kappa L_{a}}{c_{e}}-\sum_{a_{i}\in Q_{e}}\frac{L_{a_{i}}'(t)}{\bar{r}_{ie}(t)}-\text{\ensuremath{\tau}}_{\textrm{EP}}.\label{eq:final_delay_const}
\end{equation}

\vspace{-0.1cm}

The above constraint implies that to reach the desired reliability,
a maximum value of $\frac{L_{a}}{\bar{r}_{ue}(t)}$ is allowed for
the newly admitted requests to the queue of cloudlet $e$. The average
service rate $\bar{r}_{ue}(t)$ is estimated at each cloudlet $e$
for each UN $u$ within its coverage using a time-average rate estimation
method, as follows:

\vspace{-0.6cm}
\begin{equation}
\bar{r}_{ue}(t)=\nu(t)r_{ue}(t-1)+(1-\nu(t))\bar{r}_{ue}(t-1).
\end{equation}

Next, we propose a joint matching \cite{walid_matching} and caching
scheme to solve the optimization problem in (\ref{eq:objective}).

\section{Joint Task Matching and Caching\label{sec:JointTandM} }

\vspace{-0.1cm}

To simplify the computational complexity of the optimization problem
in (\ref{eq:objective}), we decouple the problem into two separate
subproblems: distributing UN tasks to cloudlets and caching popular
cacheable task results. Due to the large size of IoT networks, it
is not practical to perform task matching over the whole network set
of cloudlets and UNs. Therefore, a clustering scheme is introduced
to group UNs into disjoint sets based on spatial proximity and mutual
interest in popular tasks, followed by the calculation of a task popularity
matrix. Subsequently, a joint task distribution and caching scheme
is proposed. UN clustering and task popularity matrix calculations
are assumed to be performed during a network training period during
which information about UNs' requests and their serving cloudlets
are reported to a higher level controller, e.g. a cloud data cent\textcolor{black}{er.
While a central controller is involved in the training period calculations,
we emphasize that this process does not need to be updated as frequently
as the task distribution and caching processes, since a given user's
interests are likely to remain unchanged for a number of time instants
$N_{t}$ ($\gg1$). } %
{}

\subsection{Network Clustering and Task Popularity Matrix}

\vspace{-0.1cm}

We start by grouping UNs into $k$ disjoint clusters $\mathcal{C}_{1},\ldots,\mathcal{C}_{k}$
based on their mutual-coupling in distance and task popularity such
that a task popularity matrix, defined as $\boldsymbol{\Xi}=[\boldsymbol{\xi}_{1},\ldots,\boldsymbol{\xi}_{k}]$
is calculated, where $\boldsymbol{\xi}_{i}$ is a vector of the popularity
order of tasks in cluster $\mathcal{\mathcal{C}}_{i}$. Essentially,
identifying the similarities between neighboring UNs and their mutual
interests is the first step in bringing computing resources closer
to them. To that end, we exploit the similarity of different UNs in
terms of their similar task popularity patterns to allow cloudlets
in their proximity to store the computing results of their tasks. 

\subsubsection{Distance-based Gaussian similarity}

The Gaussian similarity metric is used to quantify the similarity
between UNs based on their inter-distance. A distance Gaussian similarity
matrix is defined as $\boldsymbol{S}_{d}=[d_{ij}]$, with $d_{ij}$
being:

\vspace{-0.2cm}

\begin{equation}
d_{ij}=\exp\left(\frac{-\parallel\boldsymbol{v}_{i}-\boldsymbol{v}_{j}\parallel^{2}}{2\sigma_{d}^{2}}\right),
\end{equation}
where $\boldsymbol{v}_{i}$ is a vector of the geographical coordinates
of UN $i$, and $\sigma_{d}$ is a similarity parameter to control
the neighborhood size.

\subsubsection{Task popularity-based similarity}

To discover the task popularity patterns of different UNs, the task
request occurrence is recorded for each UN during a training period
set. Subsequently, a task occurrence vector $\boldsymbol{n}_{u}=[n_{u,1},\ldots,n_{u,\mid\mathcal{A}_{c}\mid}]$
is calculated for each UN. This vector captures the UN's task arrival
rate and helps to build similarity between UNs. A cosine similarity
metric is considered to measure the similarity between UNs. The task
popularity similarity matrix is $\boldsymbol{S}_{p}=[p_{ij}]$, where
$p_{ij}$ is expressed as:

\vspace{-0.3cm}

\begin{equation}
p_{ij}=\frac{\boldsymbol{n}_{i}.\boldsymbol{n}_{j}}{\parallel\boldsymbol{n}_{i}\parallel\parallel\boldsymbol{n}_{j}\parallel}.
\end{equation}

\subsubsection{UN clustering and popularity matrix calculation}

Since we are interested in groups of UNs that are close to each other
and having similar task popularity patterns, we consider a similarity
matrix that blends the distance and task popularity matrices together.
The similarity matrix $S$ is calculated as:\vspace{-0.4cm}

\begin{equation}
\boldsymbol{S}=\theta\boldsymbol{S}_{d}+(1-\theta)\boldsymbol{S}_{p},\label{eq:similarity}
\end{equation}

\vspace{-0.2cm}
{\setlength{\parindent}{0cm}where $\theta$ is a parameter that adjusts
the impact of distance and task popularity. Subsequently, we use  spectral
clustering \cite{livehoods} to group UNs into $k$ disjoint clusters,
$\mathcal{C}_{1},\ldots,\mathcal{C}_{k}$. }

To bring the popular tasks closer to the network edge, the task popularity
matrix of UN clusters is reported to cloudlets so that they cache
the computing result of the most popular tasks. Accordingly, the most
preferred cluster by a cloudlet is obtained by calculating how frequently
the members of each cluster were assigned to this specific cloudlet
during the training period. The vector $\boldsymbol{\xi}_{i}$ of
tasks that are most popular for a cluster~$i$ is reported to the
cloudlets that have cluster $\mathcal{C}_{i}$ as their most preferred
cluster. The proposed UN clustering and task popularity matrix calculation
is described in Algorithm \ref{alg:clustering}.

\vspace{-0.1cm}

\begin{algorithm}[t]
\begin{algorithmic}[1]\footnotesize

\STATE  \textbf{Training phase: }For a sequence of training time
instants:
\begin{itemize}
\item Record $\boldsymbol{n}_{u}$ of each UN.
\item Calculate the similarity matrix $\boldsymbol{S}$ from (\ref{eq:similarity}).
\item Set $k_{\min}=2$ and $k_{\max}=U/2$.
\item Record the number of times a cloudlet served each UN.
\end{itemize}
\STATE  \textbf{Clustering phase:}
\begin{itemize}
\item Perform spectral clustering using the similarity matrix $\boldsymbol{S}$,
use the largest eigenvalue gap method \cite{livehoods} to select
the number of clusters $k\in\{$$k_{\min},\ldots,k_{\max}\}$.
\item Obtain $k$ disjoint clusters of UNs $\mathcal{C}_{1},\ldots,\mathcal{C}_{k}$.
\end{itemize}
\STATE \textbf{ Popularity list construction phase:}
\begin{itemize}
\item Mark a cloudlet most preferred cluster as the cluster from which it
received the highest number of requests during the training period.
\item Calculate the  task popularity matrix $\boldsymbol{\Xi}$ of each
cluster using the number of request occurrences $\boldsymbol{n}_{u}$
of its set of UNs. 
\item Report to each cloudlet the task popularity vector $\boldsymbol{\xi}_{i}$
of its most preferred cluster $\mathcal{C}_{i}$. 
\end{itemize}
\end{algorithmic}

\caption{\label{alg:clustering}UN clustering and  popularity matrix calculation. }
\end{algorithm}

\vspace{-0.1cm}

\subsection{Computing Caching Scheme }

\vspace{-0.1cm}

During network operation, cloudlets seek to minimize the service delay
of their UNs' requests by proactively caching the computing results
of the popular tasks they receive. The caches of each cloudlet are
assumed to be empty at the beginning of the network operation. As
UNs start to offload computing tasks, cloudlets will cache as many
computing results as their storage capacity allows. Once a cloudlet's
storage is full, a new arriving request that is more popular than
the least popular task currently in the cache will replace it. The
algorithm implementation per cloudlet is described in Algorithm 2. 

\begin{algorithm}[t]
\begin{algorithmic}[1]\footnotesize

\STATE  \textbf{Initialization: }
\begin{itemize}
\item Define the set $\Psi_{e}$ as the cache content of cloudlet $e$.
\item $\Psi_{e}=\phi,\quad\forall e\in\mathcal{E}$.
\end{itemize}
\STATE  \textbf{foreach} $a\in Q_{e}$ 

\STATE  \textbf{if} $\mid\Psi_{e}\mid<s_{e}$

\begin{ALC@g}\STATE  $a\rightarrow\Psi_{e}$.

\end{ALC@g}\STATE  \textbf{else if }$\mid\Psi_{e}\mid=s_{e}$

\begin{ALC@g}\STATE  \textbf{if} there exists at least one task $a_{i}\in\Psi_{e}$
with lower index than $a$ in $\xi_{e}$

\begin{ALC@g}\STATE  task $a_{i}$ is removed from $\Psi_{e}$.

\STATE  $a\rightarrow\Psi_{e}$.

\end{ALC@g}\STATE  \textbf{else}

\begin{ALC@g}\STATE  the computing result of task $a$ is not stored.

\end{ALC@g}\STATE  end\textbf{ if}

\end{ALC@g}\STATE  end\textbf{ if}

\STATE  end \textbf{foreach}

\end{algorithmic} 

\caption{\label{alg:caching}Proactive task caching algorithm. }
\end{algorithm}

Next, if the cloudlet receives a computation request of a task that
is cached in its storage, there is no need to offload the task data
or recompute the task, and only processing delay is incurred. Each
cloudlet aims to find the optimal caching policy that minimizes the
total latency.

\vspace{-0.1cm}

\subsection{UN Task Distribution}

\vspace{-0.1cm}

Our next step is to propose a task distribution scheme that solves
the constrained minimization problem in (\ref{eq:objective}). The
task distribution problem is formulated as a matching game between
UNs and cloudlets where, at each time instant, UNs requesting new
tasks are matched to a serving cloudlet aiming to minimize their service
delay. Matching theory \cite{walid_matching} is a framework that
solves combinatorial problems in which members of two sets of players
are interested in forming \emph{matching pair}s with a player from
the opposite set. Preferences of both the cloudlets and UNs, denoted
$\succ_{e}$ and $\succ_{u}$, represent how each player ranks the
players of the opposite set. 

\vspace{-0.1cm}

\begin{defn}
\label{def:matching}Given the two disjoint sets of cloudlets and
UNs $(\mathcal{E},\mathcal{U}$), a \emph{matching} is defined as
a \emph{one-to-one} mapping $\Upsilon$ from the set $\mathcal{E}\cup\mathcal{U}$
into the set of all subsets of $\mathcal{E}\cup\mathcal{U}$, such
that for each $e\in\mathcal{E}$ and $u\in\mathcal{U}$: \vspace{-0.2cm}
\end{defn}
\begin{enumerate}
\item For each $u\in\mathcal{U},$$\Upsilon(u)\in\mathcal{E}\cup u$, where
$\Upsilon(u)=u$ means that a UN is not matched to a cloudlet, but
will perform local computing instead.
\item For each $e\in\mathcal{E},$$\Upsilon(e)\in\mathcal{U}\cup\{e\}$,
where $\Upsilon(e)=e$ means that no UN is assigned to the cloudlet
$e.$
\item $\mid\Upsilon(u)\mid=1,\mid\Upsilon(e)\mid=1$;~ 4)$\Upsilon(u)=e\Leftrightarrow\Upsilon(e)=u$.
\end{enumerate}
By inspecting the problem in (\ref{eq:objective}), we can see that
the constraints (\ref{eq:match_const_1})-(\ref{eq:match_constraint_2})
are satisfied by the one-to-one mapping of the matching game. Moreover,
matching allows defining preference profiles that capture the cost
function of the players. To this end, the preference profiles of UNs
are defined so as to minimize their task service delay as follows:

\vspace{-0.3cm}

\begin{equation}
e\succ_{u}e'\Leftrightarrow D_{ea}^{\textrm{f}}(t)<D_{e'a}^{\textrm{f}}(t),\label{eq:user_preference}
\end{equation}

\vspace{-0.5cm}

\begin{equation}
u\succ_{u}e\Leftrightarrow D_{ua}^{\textrm{l}}(t)<D_{ea}^{\textrm{f}}(t).\label{eq:user_preference-1}
\end{equation}

\vspace{-0.2cm}

Note that since a UN has no information about the queue length at
each cloudlet, it considers the transmission, computing and processing
delay of its own task data in calculating its preference profile.

\begin{algorithm}[t]
\begin{algorithmic}[1]\footnotesize

\STATE  \textbf{Initialization: }all UNs and cloudlets start unmatched. 

\STATE  Each UN constructs its preference list as per (\ref{eq:user_preference})-(\ref{eq:user_preference-1}).

\STATE Each cloudlet constructs its preference list as per (\ref{eq:cloudlet_preference})-(\ref{eq:cloudlet_preference2}).

\STATE  \textbf{repeat} an unmatched UN $u$, i.e., $\Upsilon(u)=\phi$
proposes to its most preferred cloudlets $e$ that satisfies $e\succ_{u}u$.

\begin{ALC@g}\STATE  \textbf{if} $\Upsilon(e)=\phi$, 

\begin{ALC@g}\STATE UN $u$ proposal is accepted. 

\STATE $\Upsilon(e)=u$, $\Upsilon(u)=e$.

\end{ALC@g}\STATE  \textbf{elseif} $\Upsilon(e)=u'$,

\begin{ALC@g}\STATE   \textbf{if }$u'\succ_{e}u$

\begin{ALC@g}\STATE   UN $u$ proposal is rejected. 

\STATE   UN $u$ removes cloudlet $e$ from its preference list.

\end{ALC@g}\STATE  \textbf{elseif }$u\succ_{e}u'$ 

\begin{ALC@g}\STATE UN $u$ proposal is accepted. 

\STATE $\Upsilon(e)=u$,$\Upsilon(u)=e$.

\STATE   $\Upsilon(u')=\phi$. 

\STATE   UN $u'$ removes cloudlet $e$ from its preference list.

 \end{ALC@g}\STATE  \textbf{end if}

 \end{ALC@g}\STATE  \textbf{end if}

\end{ALC@g} \STATE  \textbf{until }all UNs are either matched or
not having cloudlets that satisfy $e\succ_{u}u$. in their preference
lists.\textbf{ }

 \STATE  $\Upsilon(u)=u$ for all remaining unmatched UNs.

\STATE  \textbf{Output: }a stable matching $\Upsilon$.

\end{algorithmic}

\caption{\label{alg:matching}DA algorithm for UN-cloudlet matching.}
\end{algorithm}

The utility of cloudlets will essentially reflect the latency and
reliability constraint in (\ref{eq:final_delay_const}), taking into
account the waiting time in the queue. Therefore, we define the utility
when UN $u$ is assigned to cloudlet $e$ as:

\vspace{-0.5cm}

\begin{equation}
\Phi_{eu}(t)=D_{th}\epsilon-\frac{k_{a}L_{a}}{c_{e}}-\sum_{a_{i}\in Q_{e}}\frac{L_{a_{i}}'(t)}{\bar{r}_{ie}(t)}-\text{\ensuremath{\tau}}_{\textrm{EP}}-\frac{L_{a}}{\bar{r}_{ue}(t)}.
\end{equation}

The preference of each cloudlet can be expressed as follows:

\vspace{-0.3cm}

\begin{equation}
u\succ_{e}u'\Leftrightarrow\Phi_{eu}(t)>\Phi_{eu'}(t),\label{eq:cloudlet_preference}
\end{equation}

\vspace{-0.5cm}

\begin{equation}
e\succ_{e}u\Leftrightarrow\Phi_{eu}(t)<0,\label{eq:cloudlet_preference2}
\end{equation}

\vspace{-0.2cm}
{\setlength{\parindent}{0cm}\textcolor{black}{where (\ref{eq:cloudlet_preference2})
states that a cloudlet is not interested in being matched to a UN
that will violate its reliability constraint. In other words, the
utility of each cloudlet is to seek a matching that maximizes the
difference between the right h}and side and the left hand side of
the inequality in (\ref{eq:final_delay_const}), such that the constraint
is met as a stable matching is reached. }

The above problem is a one-to-one matching gam\textcolor{black}{e.
Next, we define matching stability and provide an efficient algorithm
based on deferred acceptance (DA) \cite{walid_matching} to solve
this game.}

\vspace{-0.2cm}

\begin{defn}
\label{def:blocking_pair}Given a matching $\Upsilon$ with $\Upsilon(e)=u$
and $\Upsilon(u)=e$, and a pair $(u',e')$ with $\Upsilon(e)\neq u'$
and $\Upsilon(u)\neq e'$, $(u',e')$ is said to be blocking the matching
$\Upsilon$ and form a blocking pair if: 1) $u'\succ_{e}u$, 2) $e'\succ_{u}e$.
A matching $\Upsilon*$ is stable if there is no blocking pair.

{\setlength{\parindent}{0cm}\textbf{\emph{Remark}}\textbf{ 1}. DA
algorithm described in Algorithm \ref{alg:matching}, converges to
a two-sided stable matching of UNs to cloudlets \cite{walid_matching}.}
\end{defn}
\begin{figure}[t]
\begin{centering}
\includegraphics[bb=0bp 0bp 400bp 295bp,width=1\columnwidth]{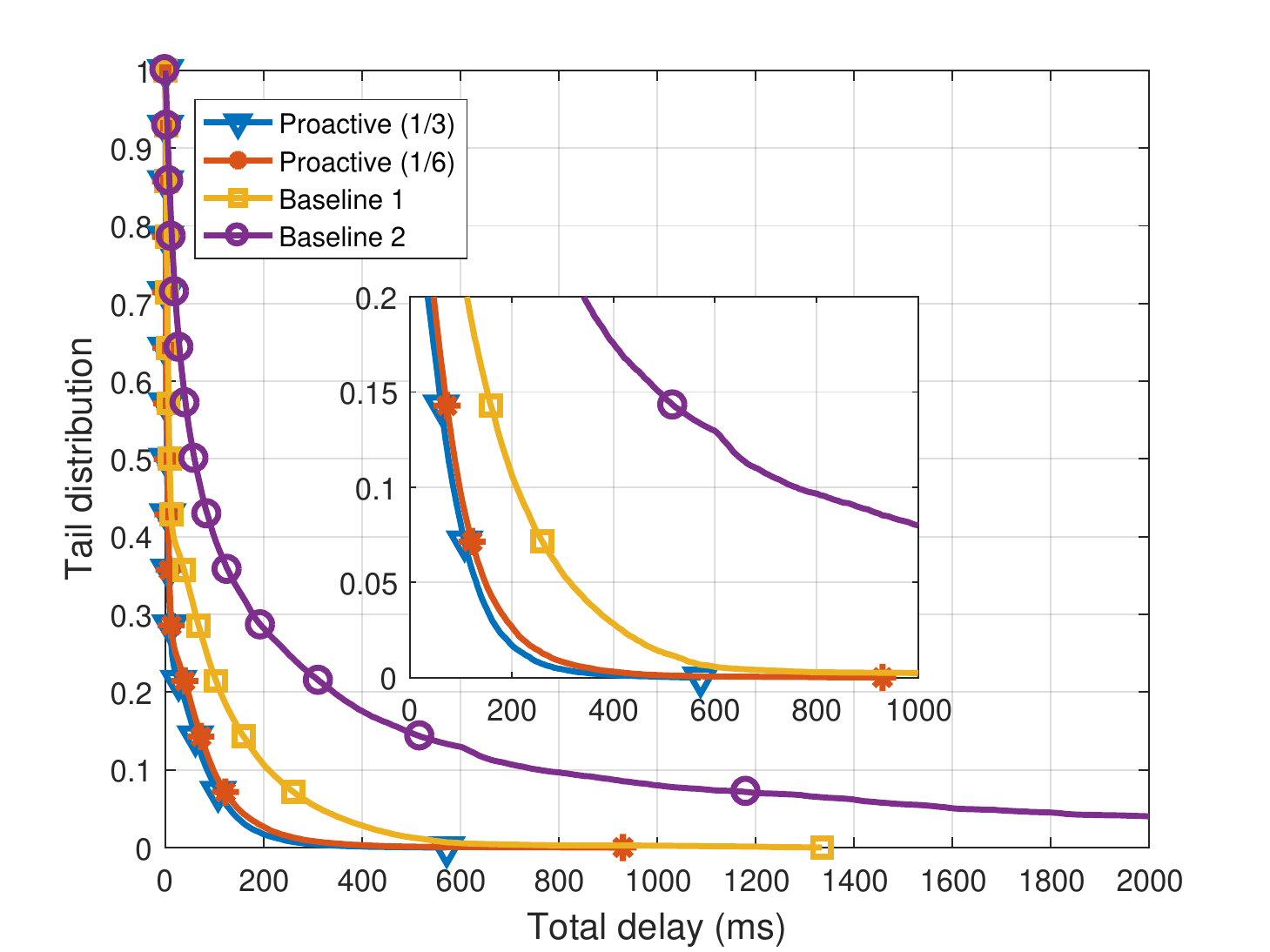}
\par\end{centering}
\centering{}\caption{\label{fig:tail_distribution}The total delay tail distribution for
the proposed (proactiveness of $1/6$ and $1/3$ of cacheable contents)
and the baseline schemes, with $E=30$ cloudlets, $U=3\times30$ UNs,
and traffic intensity of $9$ Mbps.}
\end{figure}

\begin{figure*}[tbh]
\begin{minipage}[t]{0.28\paperwidth}%
\includegraphics[bb=20bp 2bp 390bp 295bp,clip,width=0.28\paperwidth]{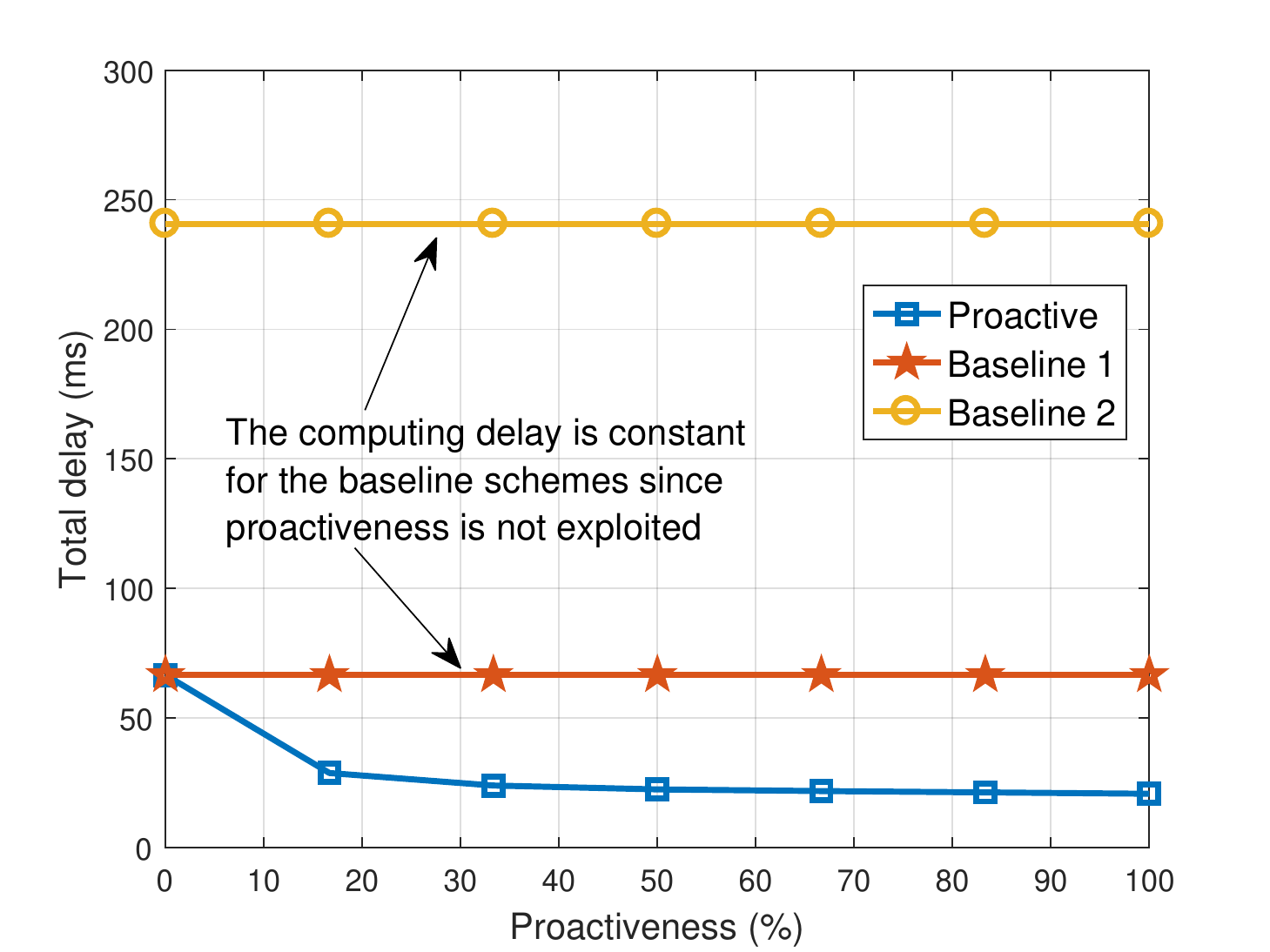}\caption{Total delay performance at different proactiveness levels\label{fig:proactiv1}}
\end{minipage}\hspace{0.05cm}%
\begin{minipage}[t]{0.29\paperwidth}%
\includegraphics[bb=23bp 2bp 413bp 295bp,clip,width=0.29\paperwidth]{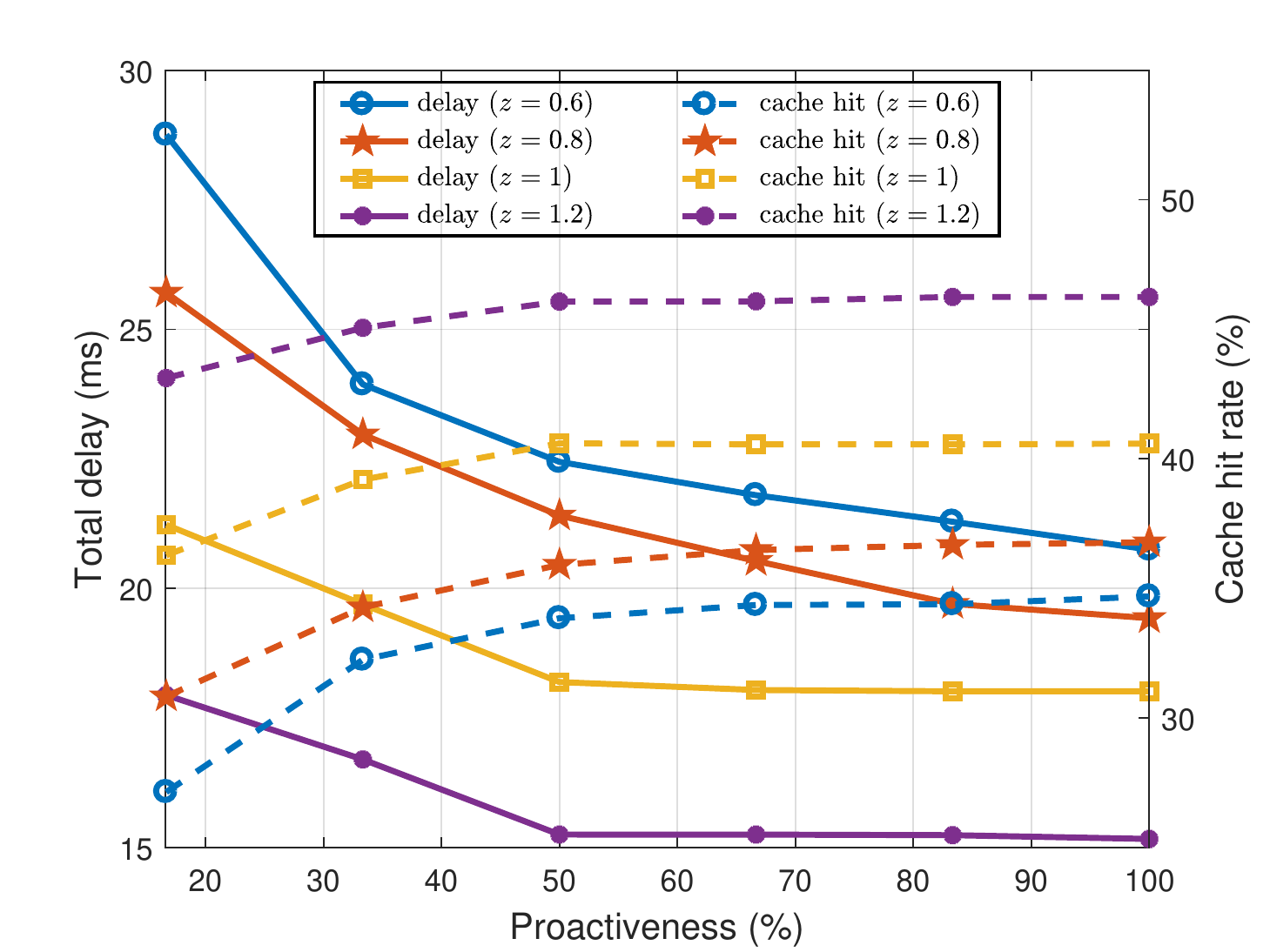}\caption{Total delay (solid lines) and cache hit rate (dashed lines) at different
values of $z$\label{fig:proactiv2}}
\end{minipage}\hspace{0.1cm}%
\begin{minipage}[t]{0.28\paperwidth}%
\includegraphics[bb=20bp 2bp 400bp 295bp,clip,width=0.28\paperwidth]{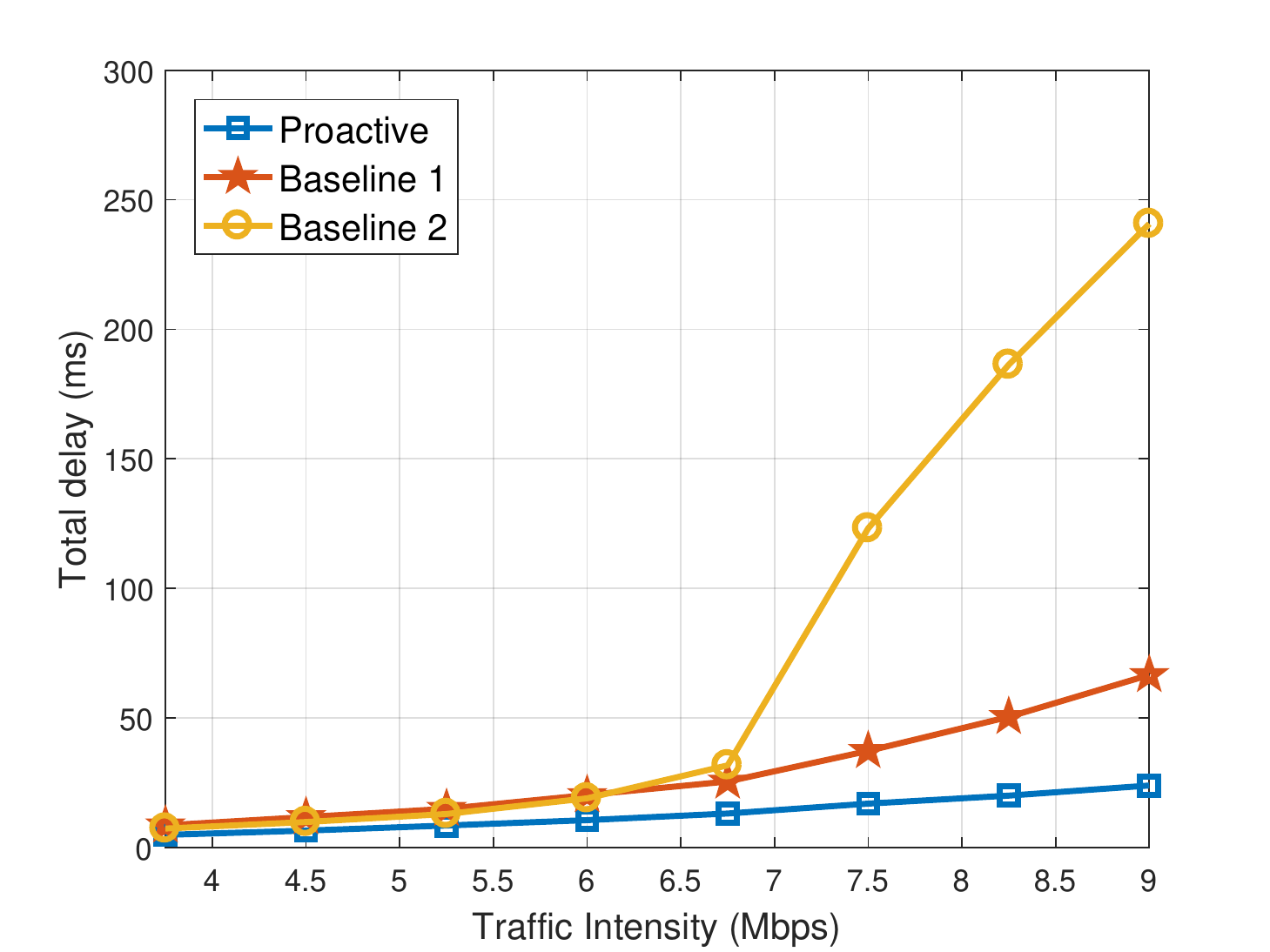}\caption{Total delay performance as the arrival traffic intensity increases\label{fig:traffic}}
\end{minipage}\\

\vspace{-0.4cm}
\end{figure*}

\vspace{-0.2cm}

\section{Simulation Results\label{sec:SimRes}}

In this section, we present and illustrate  insights from simulation
results of the proposed scheme. We also compare the proposed proactive
computing scheme against the following two baseline schemes:

\vspace{-0.1cm}

\begin{enumerate}
\item \emph{Baseline}~\emph{1}, which is a reactive version of the proposed
scheme, in which the latency and reliability constrained task distribution
scheme is considered, but with no caching capabilities in the cloudlets.
\item \emph{Baseline}~2, in which latency and reliability constraints are
not considered. Instead, UNs and cloudlets rank each other based on
the wireless access link quality, without taking delay queues or proactiveness
into account.
\end{enumerate}
We use a set of default parameters\footnote{$E=30$ cloudlets, $U=3\times30$ UNs, $\mid A\mid=90$ tasks, $z=0.6$,
$D_{\textrm{th}}=1\textrm{s}$, $\epsilon=0.01$, UN power = $20$
dBm, $\theta=0.5$, $\sigma_{d}^{2}=500$, $\nu(t)=1/t^{0.55}$, $\kappa/c_{\textrm{local}}=10^{-7},\kappa/c_{e}=10^{-8}$,
$s_{e}=10$ tasks, $\text{\ensuremath{\tau}}_{\textrm{LP}}=\textrm{Unif}(0,\tfrac{1}{8})\textrm{\,ms},$
$\text{\ensuremath{\tau}}_{\textrm{EP}}=\textrm{Unif}(\tfrac{1}{8},\tfrac{1}{4})\textrm{\,ms}$.} unless stated otherwise. Three different sets of task popularity
distributions are assigned randomly to UNs, where popularity varies
among tasks following the Zipf popularity model with parameter $z$
\cite{ejder_edge}. Accordingly, the request rate for the $i^{\textrm{th}}$
most popular task is proportional to $1/i^{z}$. Furthermore, one
third of the tasks, uniformly selected, are assumed to be cacheable.

\vspace{-0.1cm}

\subsection{Proactiveness and Computation Delay}

\vspace{-0.1cm}

In Fig.~\ref{fig:tail_distribution}, we show the tail distribution
of the instantaneous total computing delay, i.e., the complementary
cumulative distribution function (CCDF) $\bar{F_{D}}(d)=\Pr(D>d)$,
for different schemes. The proposed scheme is simulated for different
proactiveness levels of $1/3$ and $1/6$. In other words, the cloudlet
storage can store up to $1/3$ and $1/6$ of the computing results
of the cacheable tasks. From Fig.~\ref{fig:tail_distribution}, we
can see that the proposed scheme maintains a $99\%$ reliability constraint
($\epsilon=0.01$) for both the proactive and the reactive cases.
Moreover, the probability of having higher delay values significantly
decreases as the proactiveness level increases since storing more
computing results closer to UNs will further reduce the computing
delay. 

The average total delay performance is presented in Fig.~\ref{fig:proactiv1}
under different proactiveness levels. Comparing Baseline~1 and Baseline~2
schemes, about $72\%$ decrease in the average computing delay can
be seen. In Baseline~2 scheme, requests that will violate the latency
constraints are not admitted, and are computed locally instead. Furthermore,
the proactive scheme significantly decreases the computing delay as
the proactiveness level increases. By storing more computing results
close to UNs, up to $91\%$ decrease in delay is observed. 

The impact of proactiveness on the delay and cache hit rate is investigated
in Fig.~\ref{fig:proactiv2} for different discrepancy levels of
the task popularity distributions, represented by the Zipf parameter
$z$. As $z$ increases, the popularity gap between the most and least
popular tasks increases. At high values of $z$, most of the computing
requests are for the most popular tasks. Accordingly, it is possible
to serve more requests from the cache, resulting in low computing
delay and high cache hit rate, even at low proactiveness levels. On
the other hand, at low values of $z$, high proactiveness levels are
needed to store the most popular task results. Therefore, a steep
decrease in computing delay and an increase in the cache hit rate
are observed as proactiveness level increases.

\vspace{-0.1cm}

\subsection{Impact of Traffic Intensity}

\vspace{-0.1cm}

Next, we study how the performance changes with the traffic intensity.
Intuitively, at low traffic intensity conditions, cloudlets can cope
with the computing requests with minimal latency, and there is no
need to assign requests to local computing. However, at high traffic
conditions, offloading all requests causes the cloudlet queues to
grow rapidly, unless stringent latency requirements are imposed. From
Fig.~\ref{fig:traffic}, we can see that as the traffic intensity
increases, there exists a threshold point in which higher traffic
intensity will cause severe delay for the baseline scheme with unbounded
latency. Below this point, Baseline~1 scheme achieves similar or
lower delay values than Baseline~2 as there is no compelling need
to maintain latency bounds by assigning requests to local computing.
Moreover, both the reactive and proactive schemes achieve low delay
performance, with proactiveness gains of up to $65\%$ at high traffic
intensity. \vspace{-0.1cm}

\section{Conclusions\label{sec:Conc}}

\vspace{-0.1cm}

In this paper, we have proposed a task distribution and proactive
computing scheme for cache-enabled fog computing networks under ultra-reliability
and low-latency constraints. In the proposed scheme, clusters of cloudlets
and edge user nodes are formed based on spatial proximity and similar
interests in computing results. To ensure minimal computing delays,
each cluster proactively caches computing results in the storage of
its cloudlets. Moreover, we have proposed a matching algorithm to
distribute the computing tasks to cloudlets such that computing delay
is minimized and latency constraints are met. Simulation results have
shown that the proposed scheme significantly minimizes the computing
delay under different proactiveness and traffic intensity levels,
and is able to guarantee minimal latency bounds with high levels of
certainty. 

\vspace{-0.2cm}

\bibliographystyle{IEEEtran}
\bibliography{IEEEabrv,bibfile}

\end{document}